\documentclass{webofc}

\usepackage[capitalise]{cleveref}
\usepackage[varg]{txfonts}   
\usepackage{comment}

\begin{document}

\setlength{\belowcaptionskip}{-10pt}

\title{Towards the first mean pressure profile estimate with the NIKA2 Sunyaev-Zeldovich Large Program}

\author{%
  \lastname{C. Hanser}\inst{\ref{LPSC}}\fnsep\thanks{{hanser@lpsc.in2p3.fr}}
  \and  R.~Adam \inst{\ref{OCA}}
  \and  P.~Ade \inst{\ref{Cardiff}}
  \and  H.~Ajeddig \inst{\ref{CEA}}
  \and  P.~Andr\'e \inst{\ref{CEA}}
  \and  E.~Artis \inst{\ref{LPSC},\ref{Garching}}
  \and  H.~Aussel \inst{\ref{CEA}}
  \and  I.~Bartalucci \inst{\ref{Milan}}
  \and  A.~Beelen \inst{\ref{LAM}}
  \and  A.~Beno\^it \inst{\ref{Neel}}
  \and  S.~Berta \inst{\ref{IRAMF}}
  \and  L.~Bing \inst{\ref{LAM}}
  \and  O.~Bourrion \inst{\ref{LPSC}}
  \and  M.~Calvo \inst{\ref{Neel}}
  \and  A.~Catalano \inst{\ref{LPSC}}
  \and  M.~De~Petris \inst{\ref{Roma}}
  \and  F.-X.~D\'esert \inst{\ref{IPAG}}
  \and  S.~Doyle \inst{\ref{Cardiff}}
  \and  E.~F.~C.~Driessen \inst{\ref{IRAMF}}
  \and  G.~Ejlali \inst{\ref{Tehran}}
  \and  A.~Ferragamo \inst{\ref{Roma}}
  \and  A.~Gomez \inst{\ref{CAB}} 
  \and  J.~Goupy \inst{\ref{Neel}}
  \and  S.~Katsioli \inst{\ref{Athens_obs}, \ref{Athens_univ}}
  \and  F.~K\'eruzor\'e \inst{\ref{Argonne}}
  \and  C.~Kramer \inst{\ref{IRAMF}}
  \and  B.~Ladjelate \inst{\ref{IRAME}} 
  \and  G.~Lagache \inst{\ref{LAM}}
  \and  S.~Leclercq \inst{\ref{IRAMF}}
  \and  J.-F.~Lestrade \inst{\ref{LERMA}}
  \and  J.~F.~Mac\'ias-P\'erez \inst{\ref{LPSC}}
  \and  S.~C.~Madden \inst{\ref{CEA}}
  \and  A.~Maury \inst{\ref{CEA}}
  \and  P.~Mauskopf \inst{\ref{Cardiff},\ref{Arizona}}
  \and  F.~Mayet \inst{\ref{LPSC}}
  \and  A.~Monfardini \inst{\ref{Neel}}
  \and  A.~Moyer-Anin \inst{\ref{LPSC}}
  \and  M.~Mu\~noz-Echeverr\'ia \inst{\ref{LPSC}}
  \and  A.~Paliwal \inst{\ref{Roma}}
  \and  C.~Payerne \inst{\ref{LPSC}}
  \and  L.~Perotto \inst{\ref{LPSC}}
  \and  G.~Pisano \inst{\ref{Roma}}
  \and  E.~Pointecouteau \inst{\ref{Toulouse}}
  \and  N.~Ponthieu \inst{\ref{IPAG}}
  \and  G.~W.~Pratt \inst{\ref{CEA}}
  \and  V.~Rev\'eret \inst{\ref{CEA}}
  \and  A.~J.~Rigby \inst{\ref{Leeds}}
  \and  A.~Ritacco \inst{\ref{INAF}, \ref{ENS}}
  \and  C.~Romero \inst{\ref{Pennsylvanie}}
  \and  H.~Roussel \inst{\ref{IAP}}
  \and  F.~Ruppin \inst{\ref{IP2I}}
  \and  K.~Schuster \inst{\ref{IRAMF}}
  \and  A.~Sievers \inst{\ref{IRAME}}
  \and  C.~Tucker \inst{\ref{Cardiff}}
}
\institute{
  Universit\'e Grenoble Alpes, CNRS, Grenoble INP, LPSC-IN2P3, 38000 Grenoble, France
  \label{LPSC}
  \and
  Universit\'e C\^ote d'Azur, Observatoire de la C\^ote d'Azur, CNRS, Laboratoire Lagrange, France 
  \label{OCA}
  \and
  School of Physics and Astronomy, Cardiff University, CF24 3AA, UK
  \label{Cardiff}
  \and
  Universit\'e Paris-Saclay, Université Paris Cité, CEA, CNRS, AIM, 91191, Gif-sur-Yvette, France
  \label{CEA}
  \and
  INAF, IASF-Milano, Via A. Corti 12, 20133 Milano, Italy
  \label{Milan}
  \and
  Max Planck Institute for Extraterrestrial Physics, 85748 Garching, Germany
  \label{Garching}
  \and
  Aix Marseille Univ, CNRS, CNES, LAM, Marseille, France
  \label{LAM}
  \and
  Universit\'e Grenoble Alpes, CNRS, Institut N\'eel, France
  \label{Neel}
  \and
  Institut de RadioAstronomie Millim\'etrique (IRAM), Grenoble, France
  \label{IRAMF}
  \and 
  Dipartimento di Fisica, Sapienza Universit\`a di Roma, I-00185 Roma, Italy
  \label{Roma}
  \and
  Univ. Grenoble Alpes, CNRS, IPAG, 38000 Grenoble, France
  \label{IPAG}
  \and
  Institute for Research in Fundamental Sciences (IPM), Larak Garden, 19395-5531 Tehran, Iran
  \label{Tehran}
  \and
  Centro de Astrobiolog\'ia (CSIC-INTA), Torrej\'on de Ardoz, 28850 Madrid, Spain
  \label{CAB}
  \and
  National Observatory of Athens, IAASARS, GR-15236, Athens, Greece
  \label{Athens_obs}
  \and
  Faculty of Physics, University of Athens, GR-15784 Zografos, Athens, Greece
  \label{Athens_univ}
  \and
  High Energy Physics Division, Argonne National Laboratory, Lemont, IL 60439, USA
  \label{Argonne}
  \and  
  Instituto de Radioastronom\'ia Milim\'etrica (IRAM), Granada, Spain
  \label{IRAME}
  \and
  LERMA, Observatoire de Paris, PSL Research Univ., CNRS, Sorbonne Univ., UPMC, 75014 Paris, France  
  \label{LERMA}
  \and
  School of Earth \& Space and Department of Physics, Arizona State University, AZ 85287, USA
  \label{Arizona}
  \and
  IRAP, Université de Toulouse, CNRS, UPS, CNES, Toulouse, France
  \label{Toulouse}
  \and
  School of Physics and Astronomy, University of Leeds, Leeds LS2 9JT, UK
  \label{Leeds}
  \and
  INAF-Osservatorio Astronomico di Cagliari, 09047 Selargius, Italy
  \label{INAF}
  \and 
  LPENS, ENS, PSL Research Univ., CNRS, Sorbonne Univ., Universit\'e de Paris, 75005 Paris, France 
  \label{ENS}
  \and  
  Department of Physics and Astronomy, University of Pennsylvania, PA 19104, USA
  \label{Pennsylvanie}
  \and
  Institut d'Astrophysique de Paris, CNRS (UMR7095), 75014 Paris, France
  \label{IAP}
  \and
  University of Lyon, UCB Lyon 1, CNRS/IN2P3, IP2I, 69622 Villeurbanne, France
  \label{IP2I}
}

\abstract{
  High-resolution mapping of the hot gas in galaxy clusters is a key tool for cluster-based cosmological analyses. Taking advantage of the NIKA2 millimeter camera operated at the IRAM 30-m telescope, the NIKA2 SZ Large Program seeks to get a high-resolution follow-up of 38 galaxy clusters covering a wide mass range at intermediate to high redshift. The measured SZ fluxes will be essential to calibrate the SZ scaling relation and the galaxy clusters mean pressure profile, needed for the cosmological exploitation of SZ surveys. We present in this study a method to infer a mean pressure profile from cluster observations. 
  We have designed a pipeline encompassing the map-making and the thermodynamical properties estimates from maps. We then combine all the individual fits, propagating the uncertainties on integrated quantities, such as $R_{500}$ or $P_{500}$, and the intrinsic scatter coming from the deviation to the standard self-similar model. We validate the proposed method on realistic LPSZ-like cluster simulations. 
  
}
\maketitle
\section{Introduction}
\label{intro}

Galaxy clusters have been shown to provide valuable information on cosmology \cite{clustercosmo1,clustercosmo2}. Clusters are filled with a hot ionised gas that can be studied both in X-ray \cite{Sarazin88} and through the thermal Sunyaev-Zel’dovich (SZ) effect \cite{SZeffect}, a spectral distortion of the cosmic microwave background (CMB). Its magnitude is proportional to the Compton parameter $y$ which gives a measure of the electronic pressure integrated along the line of sight. According to the self similar model of structure formation we can derive an average radial pressure profile from observations scaled by mass and redshift \cite{A10,Melin2023,PACT2021}. 
This profile is a key ingredient to model the angular power spectrum of Compton parameter maps. Therefore its measurement has implications on the cosmological parameters that can be thereby inferred, \textit{e.g} $\Omega_m$, $\sigma_8$ \cite{Bolliet2018,Ruppin2019}.

Operating at the IRAM 30-m telescope with a large field of view (6.5 arcmin) and a high angular resolution ($\sim$ 17.6 and $\sim$ 11.1 arcsec) in 2 frequency bands (150 and 260 GHz respectively) \cite{NIKA2camera}, the NIKA2 camera allows us to study the intra-cluster medium in galaxy clusters with high precision. Taking advantage of the unique performance of the NIKA2 camera, the NIKA2 SZ Large Program (LPSZ \cite{LPSZ}) seeks to perform a high-resolution follow-up of 38 galaxy clusters covering a wide mass range at redshifts from $0.5$ to $0.9$ in order to re-calibrate some of the tools needed for the cosmological exploitation of SZ surveys. In particular it aims at measuring the mean pressure profile and the $Y_{SZ}$-$M$ scaling relation, thus extending previous studies in mass, redshift or angular scales ( e.g. \cite{A10,PACT2021,Planck2013PP}). 

 In this paper, we focus on the method developed to infer the mean pressure profile in the LPSZ framework. In  \cref{scanselect}, we discuss data quality assessment to control systematic effects, then in \cref{panco2} we quickly remind  the method to measure an individual cluster pressure profile. \cref{mpp} presents the method we propose to infer the mean pressure profile. We draw conclusions in Sect. 5.

\section{Map making: characterization of the data quality}
\label{scanselect}

We perform a study of the data quality in order to improve the thermodynamical properties reconstruction for all clusters. The main goal is to make a blind identification of problematic features in individual observations. We can also test the impact of low quality data on clusters' thermodynamical properties estimate.  

\begin{figure}[h]
\centering
\includegraphics[scale=0.25]{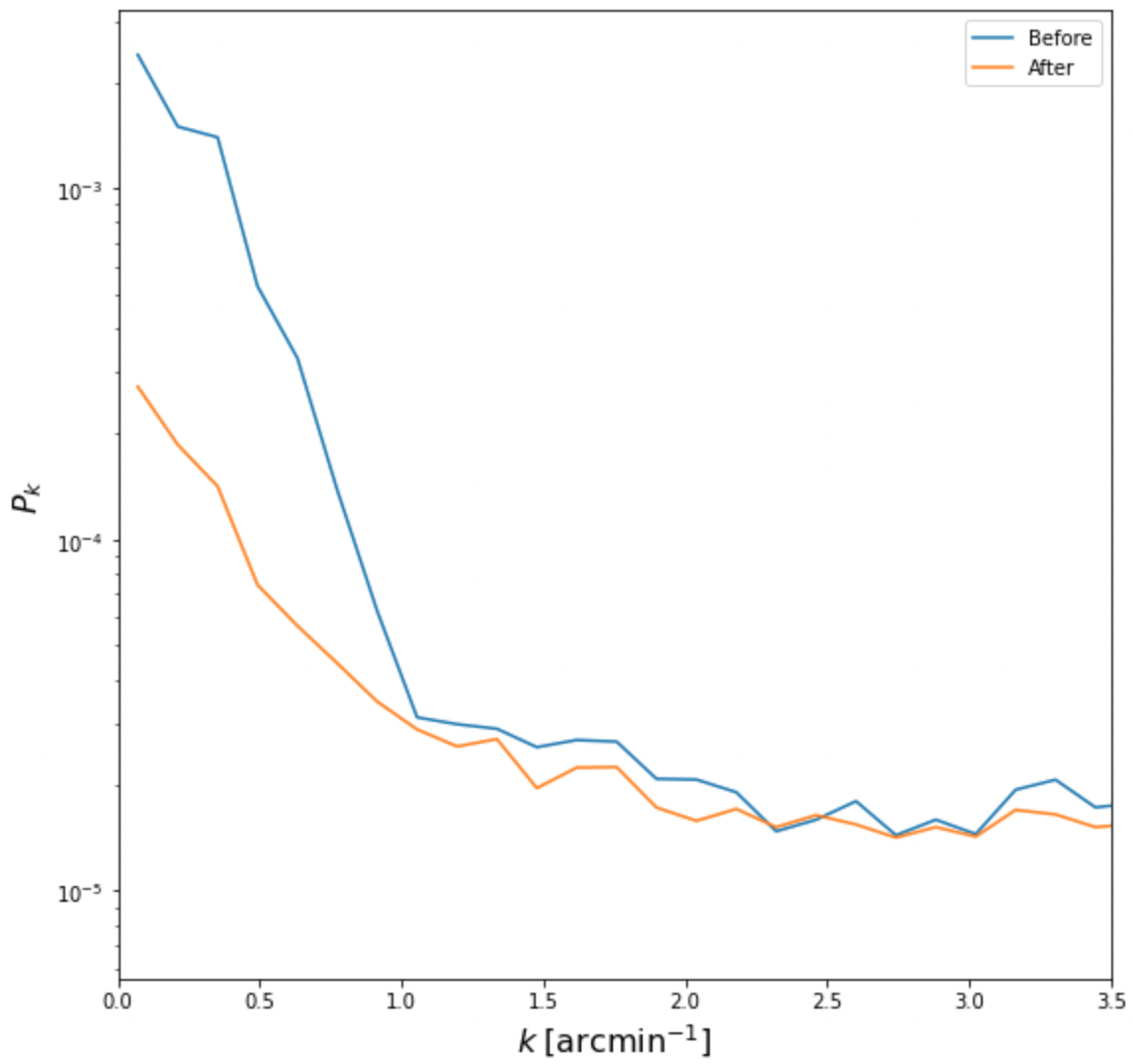}
\includegraphics[scale=0.3]{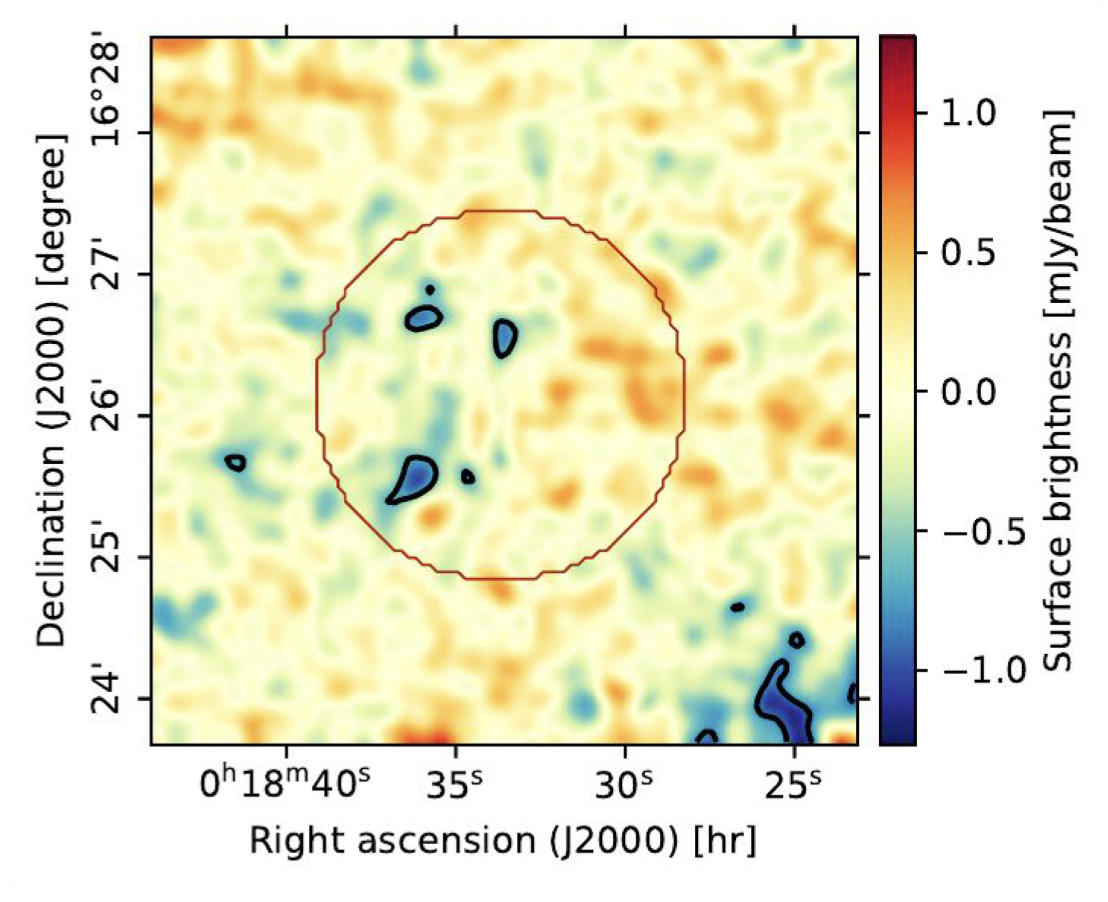}
\caption{Left panel: power spectrum of the noise maps including all observation scans (blue) and after having identified and discarded anomalous scans (orange). Right panel : NIKA2 noise map at 150 GHz obtained with the LPSZ pipeline, corresponding to the orange curve and given in mJy/beam.}
\label{Scan_sel}       
\end{figure}

This characterization is performed directly at the Time Ordered Information (TOI) level, after noise decorrelation \cite{NIKA2camera}. We defined four uncorrelated criteria for scan selection, encompassing low frequency noise properties, white noise level, integrated density flux and residual correlation between detectors. We are able to identify problematic scans (= a sample of observations) by putting a threshold on these criteria, this also allows us to detect the remaining issues in the pipeline and therefore to optimize its parameters. \cref{Scan_sel} shows on the left panel the power spectrum of the noise map before (blue) and after (orange) having applied our selection. The amplitude of the noise power spectrum after is lower than the one before since we removed a suspicious remaining noise that was present in the first iteration of the data analysis. In particular we removed two scans which were contaminating the output map. On the right panel we show the output noise map corresponding to the orange curve, we get a remaining noise compatible on average with 0. 

\section{Individual pressure profile estimate}
\label{panco2}

The 150 GHz surface brightness map of a cluster is proportional to the integrated electronic pressure along the line of sight. In order to deproject the map, we perform a Monte Carlo Markov Chain (MCMC) fit of a pressure profile model to the 150-GHz map. To do so we use an adapted version of panco2 \cite{PANCO2_pipe} similar to the version used in the first three in-depth analyses of individual LPSZ clusters \cite{Panco2_1,Panco2_2,Panco2_3}. 

Dusty galaxies and radio sources are millimeter-wavelength emitters that contribute to filling up the tSZ decrement. We thus account for their flux in the analysis as they can bias the pressure profile estimation. Other contaminants include the CMB primordial anisotropies, the CIB, the thermal diffuse emission from the galactic dust and the synchrotron radiation. Using simulation, it has been shown in \cite{Panco2_1} that all these contributions are an order of magnitude below the residual noise and are thus negligible. Moreover in order to assess the impact of the choice of a given pressure profile model, we repeat the analysis using two different models :

\begin{itemize}
    \item a generalized Navarro-Frenk-White (gNFW) model : $P_e(r) = P_0\left(\frac{r}{r_p}\right)^{-c}\left(1+\left(\frac{r}{r_p}\right)^{a}\right)^{\frac{c-b}{a}}$
    \item a radially binned model : $P_e(r_i<r<r_{i+1}) = P_i\left(\frac{r}{r_i}\right)^{-\alpha_i}$
\end{itemize}

\begin{figure}[h]
\centering
\includegraphics[scale=0.3]{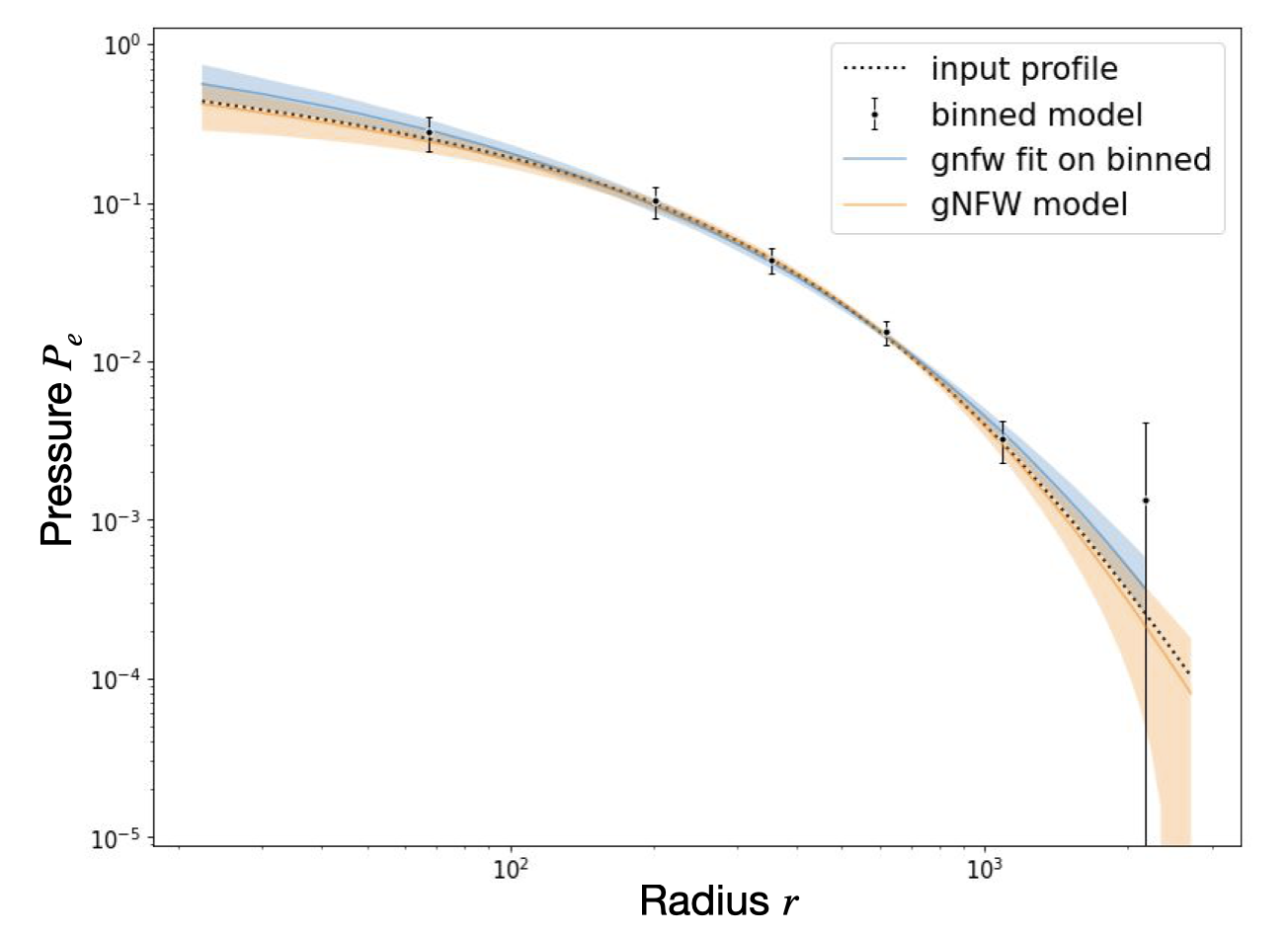}
\includegraphics[scale=0.3]{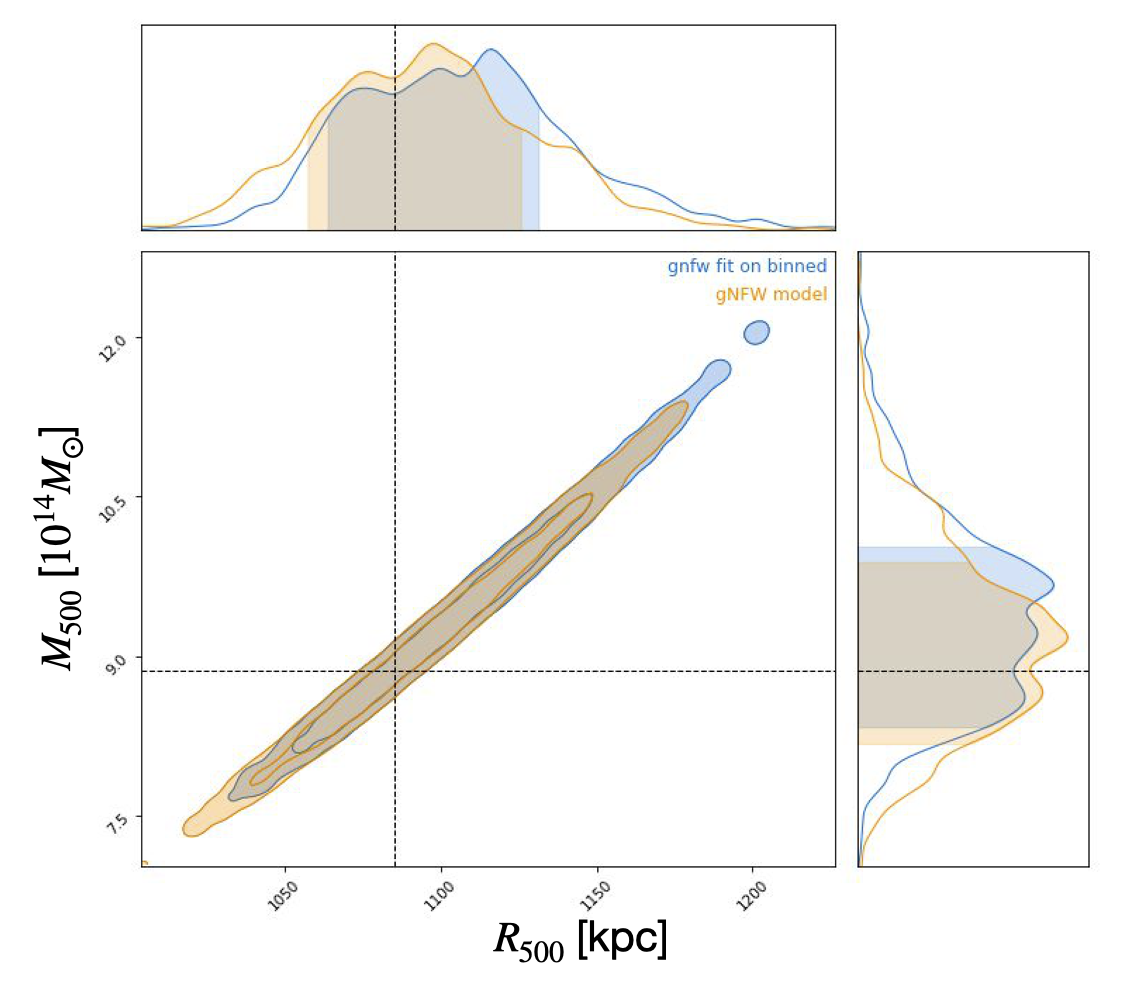}
\caption{Pressure profile and hydrostatic mass estimates on a simulated toy model cluster. Left panel: Input profile (dashed black lines) and the reconstructed pressure profiles for the binned model (black dots), the corresponding gNFW fit (blue contours) and the direct fit of the gNFW model (orange contours). Right panel:  Corresponding probability distributions of $R_{500}, M_{500}$}.
\label{v_simu}       
\end{figure}

We have tested the inference of both models using clusters realistic simulations drawn from a spherical gNFW model and covering the mass and redshift range of the LPSZ sample. These simulations also include a realistic correlated noise and the NIKA2 instrumental response. On the left panel of \cref{v_simu} we show an example of pressure profile fits using the two models for one simulated cluster. Both models recover the input profile within $1\sigma$. 

Combining the fitted spherical pressure profile with the electron density profile obtained from the XMM-Newton X-ray data, we can compute the mass profile under the hydrostatic equilibrium assumption : $M_{HSE}(<r)\propto \frac{r^2}{n_e(r)}\frac{dP_e(r)}{dr}$. We then get the probability distribution for $R_{500}, M_{500}$ knowing that $M_{500}=500\rho_{crit}4/3\pi R_{500}^3$. Since the hydrostatic mass profile depends on the derivative of the pressure profile, we also fitted a gNFW model on the binned model in order to get a continuous profile. The results of this procedure for the example simulated cluster are shown in the right panel of \cref{v_simu}. We've repeated the same analysis for every simulated clusters and have found no bias between the input and output pressure profiles. 

\section{Mean pressure profile estimate using analytical simulations}
\label{mpp}

We are now focusing on fitting the mean pressure profile (MPP) defined as $p(x) = P(r)/P_{500}$ where the pressure is normalised to a characteristic pressure $P_{500}$ \cite{A10}, and $x=r/R_{500}$. We use a normalised gNFW model with re-scaled parameters $\{p_0=\frac{P_0}{P_{500}},c_{500}=\frac{R_{500}}{r_p},\alpha,\beta,\gamma\}$. We present here the method using 45 simulated clusters with no intrinsic scatter, from which we have inferred the pressure profile with the procedure presented in \cref{panco2}. From the previous analysis we get for each cluster the gNFW fit parameters : $\{P_0, r_p,\alpha,\beta,\gamma\}$. Therefore in order to get the re-scaled parameters we need to compute $R_{500}$ and $P_{500}$.
Using the SZ+X-ray synergy we have access to $R_{500}$ (see \cref{panco2}) and we compute $P_{500}$ using Eq. (5) from \cite{A10}. The first technical point is to propagate the errors on those quantities to the combined analysis. To do so, for each cluster, we use each set of pressure profile parameters of the Markov chains to compute the corresponding $R_{500}$ and $P_{500}$ quantities and produce a Markov chain of rescaled parameters.



\begin{figure}[h]
\centering
\includegraphics[scale=0.33]{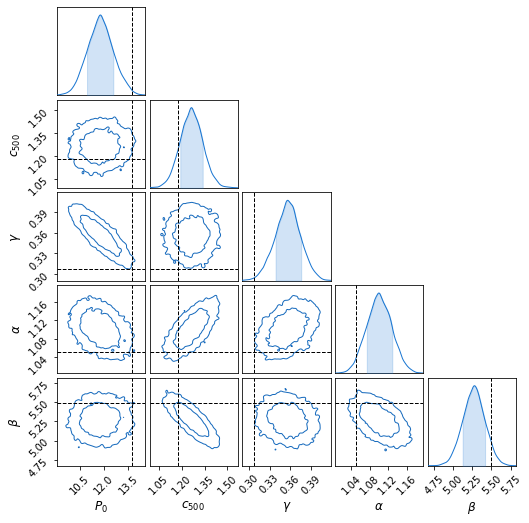}
\includegraphics[scale=0.33]{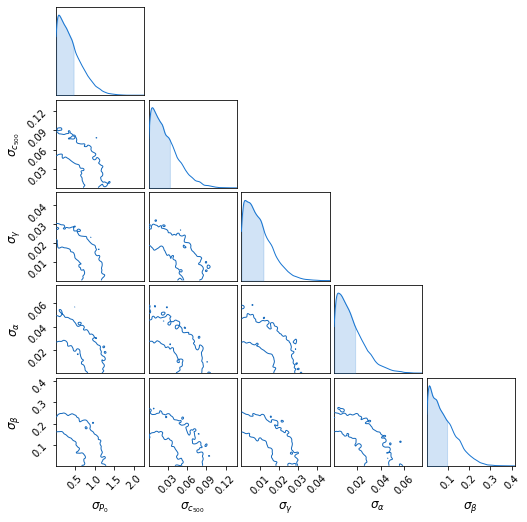}
\caption{Corner plot of the MPP parameters marginalized on the intrinsic scatter (left panel) and the corresponding intrinsic scatters (right panel). The dashed black lines show the input profile parameter values}
\label{fig:corners}       
\end{figure}

In addition to the uncertainties in the individual measurements, there is a scatter between individual re-scaled pressure profile due to the intrinsic dispersion of thermodynamical properties from a cluster to another \cite{Ghirardini2019scatter,Ghirardini2021}. To account for this effect when inferring the MPP, we adopt a Bayesian Hierarchical Modeling (see e.g. \cite{Loredo2019BHM}). Pressure profile parameters $\Vec{\theta_k}$ for the $k$-th cluster represent a measurement of the MPP parameter values affected by a scatter. Their likelihood thus follow a multivariate normal distribution of mean centered on the \textit{true} $\Vec{\theta}_{\rm MPP}$ and of covariance $\Sigma_{\rm int}$. Considering $d_k$ as the observed SZ map of the $k$-th cluster, the individual \textit{effective} cluster likelihood is given by
\begin{equation}
    \mathcal{L}_k(d_k|\Vec{\theta}_{\rm MPP}) = \int d\Vec{\theta}'\  \mathcal{L}_k(d_k|\Vec{\theta}')\ \mathcal{N}(\Vec{\theta}'|\Vec{\theta}_{\rm MPP}, \Sigma_{\rm int} ).
    \label{eq:effective likelihood}
\end{equation}
The quantity $\mathcal{L}_k(d_k|\Vec{\theta}')$ is the likelihood of the $k$-th SZ map given an arbitrary set of scaled pressure profile parameters and  $\mathcal{N}(\Vec{\theta}'|\Vec{\theta}_{\rm MPP}, \Sigma_{\rm int} )$ is the probability of fitting this arbitrary parameter set $\Vec{\theta}'$ given the true MPP parameters and their intrinsic covariance. We approximate the mapping of each individual likelihood over the parameter space by a Gaussian function\footnote{We first map the function $f_k(\vec{\theta})=\mathcal{L}_k(d_k|\Vec{\theta})$ over the parameter space using a MCMC sampler. We derive the corresponding mean $\langle \theta \rangle_k$ and parameter covariance $\mathcal{C}_k$ from the Markov chains to get $f_k(\vec{\theta})\propto \mathcal{N}(\Vec{\theta}|\langle \theta \rangle_k, \mathcal{C}_k )$.}, that enables us to compute the complex multidimensional integral in \cref{eq:effective likelihood} with little effort \cite{Duda2018MvariateGaussian}. 

Second, we consider a simplified approach for the intrinsic covariance $\Sigma_{\rm int}$ between the parameters $\{p_0, c_{500},\alpha,\beta,\gamma\}$ by neglecting the correlations between them\footnote{These correlations can be investigated using realistic hydro-dynamical simulations.}, we then only account for the intrinsic scatter of each parameter of the model. The full likelihood is the product of each effective likelihood as given in \cref{eq:effective likelihood} over the cluster sample. We use a MCMC algorithm to draw the posterior distribution of the MPP parameters and the corresponding intrinsic scatters. 


We show in blue in \cref{fig:corners} the 1 and 2$\sigma$ contours of the MPP parameters posteriors (left panel), marginalized over the intrinsic scatters $\sigma_{\rm int, \alpha}^2=(\Sigma_{\rm int})_{\alpha\alpha}$. We recover the input profile  within $2\sigma$, although we see some bias along the known $\gamma-p_0$ degeneracy \cite{Nagai2007}. We show each scatter posterior (right panel) $\sigma_{\rm int, \alpha}$ which peaks at $0$ and decreases for higher positive values, which validates our modeling input since the simulated cluster pressure profiles were generated without intrinsic noise. Accordingly, we find the same uncertainties on the MPP parameters as those obtained while imposing a fixed null-value to all intrinsic scatter parameters. However, with a realistic dataset such as the LPSZ cluster sample we expect the non-zero scatter of cluster intrinsic properties to increase the uncertainties of the MPP estimate. 

\section{Conclusion}

In this work we show how to improve NIKA2 cluster maps by characterizing the data quality and thus further improve the pipeline. We then show how we compute clusters thermodynamical properties from these maps and we validated the methods on realistic LPSZ simulations. Finally we propose a new method to compute a mean pressure profile for galaxy clusters using all individual information. This method features the propagation of uncertainties on $R_{500}$ and $P_{500}$ on each cluster and takes into account the intrinsic scatter due to deviation from the self-similar model. We validate this method on realistic simulations but with no intrinsic noise. As a conclusion we have developed a robust comprehensive pipeline to infer the first mean pressure profile of the NIKA2-LPSZ cluster sample. This upcoming measurement is anticipated to improve the accuracy of the SZ-based cluster science.

\section*{Ackowledgements}
\begin{small}
We would like to thank the IRAM staff for their support during the observation campaigns. The NIKA2 dilution cryostat has been designed and built at the Institut N\'eel. In particular, we acknowledge the crucial contribution of the Cryogenics Group, and in particular Gregory Garde, Henri Rodenas, Jean-Paul Leggeri, Philippe Camus. This work has been partially funded by the Foundation Nanoscience Grenoble and the LabEx FOCUS ANR-11-LABX-0013. This work is supported by the French National Research Agency under the contracts "MKIDS", "NIKA" and ANR-15-CE31-0017 and in the framework of the "Investissements d’avenir” program (ANR-15-IDEX-02). This work has benefited from the support of the European Research Council Advanced Grant ORISTARS under the European Union's Seventh Framework Programme (Grant Agreement no. 291294). E. A. acknowledges funding from the French Programme d’investissements d’avenir through the Enigmass Labex. A. R. acknowledges financial support from the Italian Ministry of University and Research - Project Proposal CIR01$\_00010$. S. Katsioli acknowledges support provided by the Hellenic Foundation for Research and Innovation (HFRI) under the 3rd Call for HFRI PhD Fellowships (Fellowship Number: 5357).
\end{small}

\end{document}